\documentclass[aps,twocolumn,pra,groupedaddress,superscriptaddress,nofootinbib]{revtex4}
\usepackage{graphicx,amsmath,amssymb,amsbsy,subfigure,hyperref,bbm,times,txfonts,float}
\usepackage{bbold}
\hypersetup{ colorlinks=true,linkcolor=blue}
\graphicspath{{figure/}}
\begin{document}

\newcommand{\ket}[1]{|#1\rangle}
\newcommand{\bra}[1]{\langle#1|}
\newcommand{\ketbra}[1]{| #1\rangle\!\langle #1 |}
\newcommand{\kebra}[2]{| #1\rangle\!\langle #2 |}
\newcommand{\id}{\mathbbm{1}}
\newcommand{\ohm}{\Omega_{\rm CQ}}
\newcommand{\rhobd}{\rho^{\vec{c}}_{AB}}
\providecommand{\tr}[1]{\text{tr}\left[#1\right]}
\providecommand{\tra}[1]{\text{tr}_A\left[#1\right]}
\providecommand{\trb}[1]{\text{tr}_B\left[#1\right]}
\providecommand{\abs}[1]{\left|#1\right|}
\providecommand{\sprod}[2]{\langle#1|#2\rangle}
\providecommand{\expect}[2]{\bra{#2} #1 \ket{#2}}

\title{Quantumness and memory of one qubit in a dissipative cavity under classical control}

\author{Hossein Gholipour}
\affiliation{Department of Physics, University of Guilan, P. O. Box 41335--1914, Rasht, Iran}

\author{Ali Mortezapour}
\email{mortezapour@guilan.ac.ir}
\affiliation{Department of Physics, University of Guilan, P. O. Box 41335--1914, Rasht, Iran}

\author{Farzam Nosrati}
\affiliation{Dipartimento di Ingegneria, Universit\`{a} di Palermo, Viale delle Scienze, Edificio 9, 90128 Palermo, Italy}
\affiliation{INRS-EMT, 1650 Boulevard Lionel-Boulet, Varennes, Québec J3X 1S2, Canada}

\author{Rosario Lo Franco}
\email{rosario.lofranco@unipa.it}
\affiliation{Dipartimento di Ingegneria, Universit\`{a} di Palermo, Viale delle Scienze, Edificio 6, 90128 Palermo, Italy}
\affiliation{Dipartimento di Fisica e Chimica, Universit\`a di Palermo, via Archirafi 36, 90123 Palermo, Italy}

\begin{abstract}

Hybrid quantum-classical systems constitute a promising architecture for useful control strategies of quantum systems by means of a classical device. Here we provide a comprehensive study of the dynamics of various manifestations of quantumness with memory effects, identified by non-Markovianity, for a qubit controlled by a classical field and embedded in a leaky cavity. We consider both Leggett-Garg inequality and  quantum witness as experimentally-friendly indicators of quantumness, also studying the geometric phase of the evolved (noisy) quantum state. We show that, under resonant qubit-classical field interaction, a stronger coupling to the classical control leads to enhancement of quantumness despite a disappearance of non-Markovianity. Differently, increasing the qubit-field detuning (out-of-resonance) reduces the nonclassical behavior of the qubit while recovering non-Markovian features. We then find that the qubit geometric phase can be remarkably preserved irrespective of the cavity spectral width via strong coupling to the classical field. 
The controllable interaction with the classical field inhibits the effective time-dependent decay rate of the open qubit. These results supply practical insights towards a classical harnessing of quantum properties in a quantum information scenario.      

\end{abstract}

\date{\today}


\maketitle

\section{Introduction}

Quantum features, such as coherence, entanglement, discord and nonlocality, have been introduced as physical resources \cite {RevModPhys.89.041003,RevModPhys.86.419,horodecki2009quantum,RevModPhys.84.1655,LFCSciRep,sciarrinolofranco2017,IndPRL2018,castellini2018} for performing special tasks, for instance quantum computing \cite{LG1}, quantum teleportation \cite{LG2} or quantum dense coding \cite{LG3}, which are not classically plausible \cite{LG4}. Hence, nowadays, identifying and controlling quantumness of the systems is of great importance. A common procedure for characterizing non-classicality is to test criteria which are constructed based on classical constraints. Obviously, violation of these criteria will reveal quantumness of the systems spontaneously \cite{RevModPhys.86.419}. For instance, Leggett-Garg inequality, which is based on the classical notions of macroscopic realism and noninvasive measurability, has been proposed to detect quantum behavior of a single system \cite{LG5,LG6}. It is noteworthy that the assumption of macroscopic realism indicates that macroscopic systems remain in a defined state with well-defined pre-existing value at all times, while noninvasive measurability imply that it is possible to measure, in principle, this pre-existing value without perturbing the subsequent dynamics of system. In this regard, a violation of the Leggett-Garg inequality will disclose the existence of non-classical temporal correlations in the dynamics of an individual system \cite{LG5,LG6}. This has led to a substantial amount of literature attempting to characterize system non-classicality by Leggett-Garg inequalities both theoretically \cite{LG7,LG8,LG9,LG10,LG11} and experimentally \cite{LG12,LG13}. Along this direction, robust quantum witnesses have been then introduced as alternative to the Leggett-Garg inequality, which may help in reducing the experimental requirements in verifying quantum coherence in complex systems \cite{QWNori,PhysRevA.87.052115,schild2015maximum}.

As another peculiar trait, a quantum system which evolves cyclically can acquire a memory of its evolution in the form of a geometric phase (GP). The geometric phase was first introduced in optics by Pancharatnam while he was working on polarized light \cite{GP1}. Successively, Berry theoretically explored GP in a closed quantum system which undergoes adiabatic and cyclic evolution \cite{GP2}. Afterwards, GP has been extended to systems under non-adiabatic cyclic and non-adiabatic non-cyclic evolutions \cite{GP3,GP4, GP5,GP6,GP7,GP8,GP9,GP10,GP11}. Owing to the fact that system-environment interactions play an important role for the realization of some specific operations, the study of the geometric phase has been also applied to open quantum systems. Along this route, some works have analyzed the correction to the GP under the influence of an external environment using different approaches \cite{GP12,GP13,GP14,GP15,GP16,GP17,GP18,GP19,GP20,GP21,GP22,GP23,GP24,GP25,GP26,GP27}. The global properties of the GP make it appropriate for constructing quantum gates with highest fidelities with applications within the so-called geometric quantum computation \cite{GP28,GP29,GP30,GP31,GP32,GPQI}. 

On the other hand, the dynamics of open quantum systems is categorized into two different classes of dynamical processes known as Markovian (memoryless) and non-Markovian (memory-keeping) regimes \cite{NM1,NM2}. An open quantum system undergoing Markovian dynamics, irreversibly transfers information to its surrounding environment leading to the information erasure. Differently, non-Markovian dynamics encompasses memory effects allowing a partial recovery of information of the quantum system, even after complete disappearance \cite{lofrancoreview,NM2,NM3,aolitaReview,CarusoReview}. Such a feature temporarily counteracts the detrimental effect of the surrounding environment. A plenty of suitable measures have appeared in the literature in an attempt to quantify non-Markovianity in quantum systems \cite{NM3,NM4,NM5,NM6,NM7,NM8,NM9,NM10,NM11}, also identifying it as a potential resource \cite{NM14,NM15}. Recently, effects of frequency modulation \cite{NM12}, qubit velocity \cite{NM13,mortezapour2018,mortezapour2017} and weak measurement \cite{NM14} on the non-Markovianity of a qubit inside leaky cavities have been studied. 

Memory effects alone are not usually sufficient to permit the desired long-time preservation of the quantum features of an open system. External control is thus required to this scope. One of the most intriguing aspects within this context is given by the possibility to harness quantumness by classical fields in hybrid quantum-classical systems \cite{Milburn4469,Altafini}. Such hybrid systems have in fact a fundamental role for the comprehension of the effects of classical environments on quantum features and can then constitute convenient platforms for their protection against the noise \cite{Calvani23042013,lofranco2012PRA,LoFrancoNatCom,adeline2014,PhysRevA.95.052126,trapani2015,benedetti2013,rossiparis,benedettiIJQI,PhysRevA.87.042310,bordone2012,ranganiAOP,benedettiParisIJQI,trapani2016,cresserOptComm,
darrigo2012AOP,LeggioPRA,darrigo2014IJQI,darrigo2013hidden,
bellomo2012noisylaser,lofranco2012PS}.   

To enlarge our knowledge about classical control of an open quantum system, here we focus on a qubit inside a cavity subject to a classical driving field. It is worth recalling that entanglement dynamics \cite{class1}, quantum Fisher information \cite{class2}, quantum speed-up \cite{class3}, coherence dynamics \cite{class4} and non-Markovianity \cite{class3} of such a model have been studied, the latter limited to the resonant interaction of the classical field with the qubit. 
In this work, we aim at providing insights concerning most fundamental quantum traits of the system, which can be directly measured by feasible experiments and also useful in a quantum information scenario. In particular, we investigate how non-classicality, identified by both Leggett-Garg inequality and quantum witness, geometric phase and non-Markovianity are influenced, under quite general conditions, by the adjustable classical field. We extend our study to the non-resonant interaction and compare the results with those obtained for the resonant case.

The paper is organized as follows. In Sec.~\ref{secBe11} we review the model \cite{class1,class2,haikka2010}, giving the explicit expression of the evolved reduced density matrix. In Sec.~\ref{secDM}, using Leggett-Garg inequalities, we show the time behavior of non-classicality (quantumness) of the system. The results regarding GP and non-Markovianity are presented in Sec.~\ref{secU} and Sec.~\ref{secNon}, respectively. Finally, the dynamics of quantum witness is given in Sec.~\ref{secQW}. In Sec.~\ref{secCon} we summarize our conclusions.

\section{The system}\label{secBe11}

\begin{figure}[t!]
\begin{center}
\includegraphics[width=0.45\textwidth]{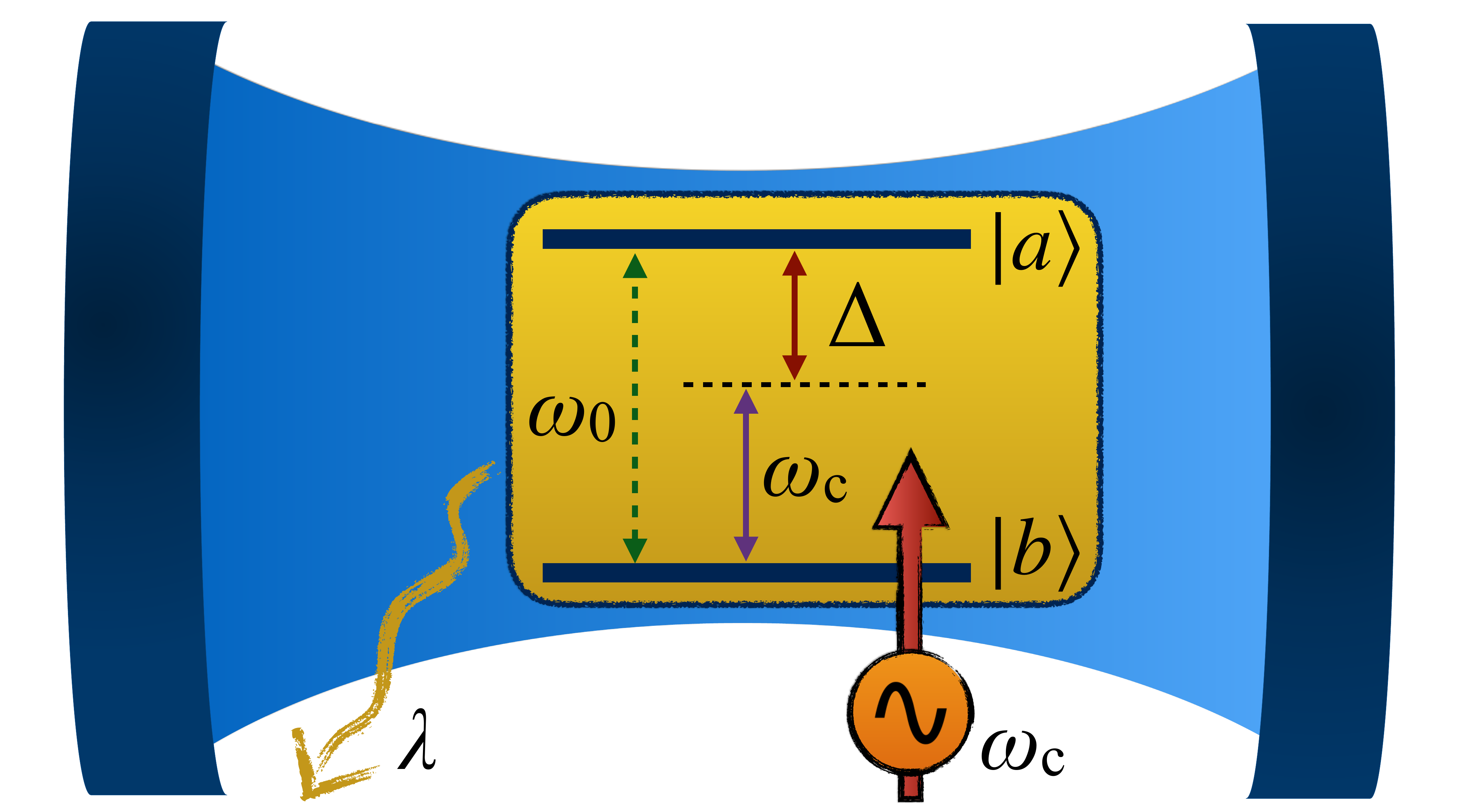}
\end{center}
\caption{\textbf{Sketch of the system.} A two-level emitter (qubit) with transition frequency $\omega_0$ is embedded in a high-$Q$ cavity with photon losses, having a spectral width (decay rate) $\lambda$. The qubit is controlled by a classical field of frequency $\omega_c$, which can be resonant or out-of-resonance with $\omega_0$, having detuning $\Delta$.}
\label{AR1}
\end{figure}

We consider a qubit (two-level emitter) of excited and ground states $\ket{a}$ and $\ket{b}$, respectively, with transition frequency $\omega_{0}$, driven by an external classical field of frequency $\omega_{c}$ and embedded in a zero-temperature reservoir formed by the quantized modes of a high-$Q$ cavity, as depicted in Fig.~\ref{AR1}. Under the dipole and rotating-wave approximations (detuning $\Delta=\omega_{0}-\omega_{c}\ll \omega_{0}, \omega_{c}$), the associated Hamiltonian of the system can be written as ($\hbar=1$) \cite{class1,class2}
\begin{equation} 
\label{eq:1}  
\hat{H}=\omega_{0}\hat{\chi}_z/2+\sum_{k}\omega_{k}\hat{a}^{\dagger}_{k}\hat{a}_{k}+\lbrace\Omega e^{-i\omega_{c}t}\hat{\chi}_{+}+\sum_{k}g_{k}\hat{\chi}_{+}\hat{a}_{k}+c.c\rbrace,
\end{equation}
where $\omega_{k}$ are the frequencies of the cavity quantized mode $k$, $\hat{\chi}_z=\ket{a}\bra{a}-\ket{b}\bra{b}$, $\hat{\chi}_{+}=\ket{a} \bra{b}$ ($\hat{\chi}_{-}=| b\rangle \langle a|$) denotes the qubit raising (lowering) operator, while $\hat{a}_{k}$ ($\hat{a}^{\dagger}_{k}$) is the annihilation (creation) operator of the $k$-th cavity mode. In addition, $\Omega$ and $g_{k}$ represent the coupling strengths of the interactions of the qubit with the classical driving field and with the cavity modes, respectively. We assume that $\Omega$ is small compared to the atomic and laser frequencies ($\Omega\ll\omega_{0},\omega_{c}$).
Using the unitary transformation $U=e^{-i\omega_{c}\hat{\chi}_z t/2}$, the Hamiltonian of the system in the rotating reference frame becomes \cite{class1,class2}
\setlength\arraycolsep{1.4pt}\begin{eqnarray}
\label{eq:2}
\hat{H}_\mathrm{eff}&=&\hat{H}_{I} +\hat{H}_{II},\nonumber\\
\hat{H}_{I}&=& \Delta\hat{\chi}_z/2+\lbrace\Omega\hat{\chi}_{+}+c.c\rbrace, \nonumber\\
\hat{H}_{II}&=& \sum_{k}\omega_{k}\hat{a}^{\dagger}_{k}\hat{a}_{k}+\sum_{k}\lbrace g_{k}\hat{\chi}_{+}\hat{a}_{k}e^{i\omega_{c}t}+c.c\rbrace,
\end{eqnarray} 
where $\Delta$ is the detuning between qubit transition frequency and classical field frequency defined above. 
By introducing the dressed states
 \setlength\arraycolsep{1.4pt}\begin{eqnarray}
\label{eq:3}
| A\rangle =\sin(\eta/2) | b\rangle + \cos(\eta/2)| a\rangle, \quad
| B\rangle =\cos(\eta/2)| b\rangle - \sin(\eta/2)| a\rangle,
\end{eqnarray} 
which are the eigenstates of $\hat{H}_{I}$, the effective Hamiltonian can be rewritten as \cite{class1,class2,shen2014exact,liu2006scalable}
 \setlength\arraycolsep{1.4pt}\begin{eqnarray}
\label{eq:4} 
\hat{H}^{\prime}_\mathrm{eff} &=& \omega_{D}\hat{\sigma}_z/2+\sum_{k}\omega_{k}\hat{a}^{\dagger}_{k}\hat{a}_{k}
+\cos^{2}(\eta/2)\sum_{k}\lbrace g_{k}\hat{\sigma}_{+}\hat{a}_{k}e^{i\omega_{c}t}+c.c \rbrace,
\end{eqnarray}  
where $\hat{\sigma}_z=\ket{A}\bra{A}-\ket{B}\bra{B}$, $\eta=\arctan(2\Omega/\Delta)$ and $\omega_{D}=\sqrt{\Delta^{2}+4\Omega^{2}}$ denotes the dressed qubit frequency. Moreover, 
$\hat{\sigma}_{-}=| B\rangle \langle A|$ ($\hat{\sigma}_{+}=| A\rangle \langle B|$) represents the new lowering (raising) qubit operator. Notice that the dressed states of Eq. (\ref{eq:3}) are just two linear combinations of the qubit bare states $\ket{a}$ and $\ket{b}$, which are conveniently introduced being the eigenvalues of the Hamiltonian $\hat{H}_{I}$ related to the interaction between the qubit and the external classical field \cite{class2}. It is noteworthy that the validity of the effective Hamiltonian of Eq. \eqref{eq:4} stems both from the rotating wave approximation ($\Delta, \Omega\ll \omega_0, \omega_c$) and from neglecting the arising Lamb shift term (proportioinal to $\hat{\sigma}_z$), which produces a small shift in the energy of the qubit with no qualitative effects on its dynamics \cite{class2,haikka2010}. In the following we shall focus on detectable qualitative behaviors of some observable quantum properties of the driven qubit and all the parameters are taken so to fulfill the above approximations.

We assume the overall system to be initially in a product state with the qubit in a given coherent superposition of its states ($\cos\theta| A\rangle+\sin\theta| B\rangle $) and the reservoir modes in the vacuum state ($| 0\rangle$), that is
\begin{equation} 
\label{eq:5}  
| \Psi(0)\rangle =(\cos\theta\ | A\rangle + \sin\theta\ | B\rangle) 
| 0\rangle.
\end{equation}
Hence, the evolved state vector of the system at time $t$ is
\setlength\arraycolsep{1.4pt}\begin{eqnarray}
\label{eq:6} 
| \Psi(t)\rangle = A(t) \cos\theta\ | A\rangle| 0\rangle+\sin\theta\ | B\rangle| 0\rangle 
+\sum_{k}B_{k}(t)| B\rangle| 1_{k}\rangle,
\end{eqnarray}
where $| 1_{k}\rangle$ represents the cavity state with a single photon in mode $k$ and $B_{k}(t)$ is the corresponding probability amplitude.
The evolution of the state vector obeys the Schr\"{o}dinger equation. Substituting Eq.~(\ref{eq:6}) into the Schr\"{o}dinger equation, we obtain the integro-differential equation for $A(t)$ as
\begin{equation} 
\label{eq:7}  
\dot{A}(t)+ \cos^{4}(\eta/2)\int_{0}^{t}dt^{\prime}F(t,t^{\prime})A(t^{\prime})= 0,
\end{equation}
where the kernel $F(t,t^{\prime})$ is the correlation function defined in terms of continuous limits of the environment frequency
\begin{equation} 
\label{eq:8}  
F(t,t^{\prime})=\int_{0}^{\infty}J(\omega_{k})e^{i(\omega_{D}+\omega_{c}-\omega_{k})(t-t^{\prime})}d\omega_{k}.
\end{equation}
Here, $J(\omega_{k})$ represents the spectral density of reservoir modes. We choose a Lorentzian spectral density, which is typical of a structured cavity \cite{NM1}, whose form is
\begin{equation} 
\label{eq:9}  
J(\omega_{k})=\frac{1}{2\pi}\frac{\gamma\lambda^{2}}{[ (\omega_{0}-\omega_{k}-\delta)^{2}+\lambda^{2}]} ,
\end{equation}
where $\delta=\omega_{0}-\omega_{n}$ denotes the detuning between the center frequency of the cavity modes and $\omega_{0}$. The parameter $\gamma$ is related to the microscopic system-reservoir coupling constant (qubit decay rate), while $\lambda$ defines the spectral width of the cavity modes. It is noteworthy that the parameters $\gamma$ and $\lambda$ are related, respectively, to the reservoir correlation time $\tau_{r}$ and to the qubit relaxation time $\tau_{q}$, as $\tau_{r}=\lambda^{-1}$ and $\tau_{q}\approx\gamma^{-1}$ \cite{NM1}. It is noteworthy that the reservoir correlation time $\tau_{r}$ and the qubit relaxation time $\tau_{q}$ the are related to parameters $\gamma$ and $\lambda$ as $\tau_{r}=\lambda^{-1}$ and $\tau_{q}\approx\gamma^{-1}$ respectively \cite{NM1}. Qubit-cavity weak coupling occurs for $\lambda>\gamma$ $(\tau_{r}<\tau_{q})$; the opposite condition $\lambda<\gamma$ $(\tau_{r}>\tau_{q})$ thus identifies strong coupling. The larger the cavity quality factor $Q$, the smaller the spectral width $\lambda$ and so the photon decay rate. To remain within the approximations of the Hamiltonian model, we shall choose $\lambda<\gamma$ \cite{haikka2010}. 

With the Lorentzian spectral density $J(\omega_{k})$ above, the kernel of Eq.~(\ref{eq:8}) becomes
\begin{equation} 
\label{eq:10}  
F(t,t^{\prime})=(\gamma\lambda/2)e^{-M(t-t^{\prime})},
\end{equation}
with $M=\lambda-i(\omega_{D}+\delta-\Delta)$. Substituting the resulting kernel into Eq.~(\ref{eq:7}) yields   
\begin{equation} \label{eq:11}  
A(t)=e^{-Mt/2}\left\lbrace\cosh( Ft/4)+\frac{2M}{F}\sinh ( Ft/4) \right\rbrace,
\end{equation}
where $F=\sqrt{4M^{2}-2\gamma\lambda(1+\cos\eta)^{2}}$ (the expression of the amplitude $B_k(t)$ of Eq. \eqref{eq:6} is reported in Appendix \ref{amplitudeB}). 

The time-dependent reduced density matrix of the qubit is obtained, in the dressed basis $\{ | A\rangle, | B \rangle \} $, by tracing out the cavity degrees of freedom from the evolved state of Eq.~(\ref{eq:6}), that gives
 \begin{equation} 
\label{eq:13}  
\rho(t)=\left(\begin{array}{cc} \cos^{2}\theta \ \vert A(t)\vert^{2}  &  \frac{1}{2}\sin(2\theta)\ A(t)  \\ 
 \frac{1}{2}\sin(2\theta)\ A^{*}(t) &  1-\cos^{2}\theta \ \vert A(t)\vert^{2}\\ 
\end{array}
\right).
\end{equation}                                       
Having this expression of $\rho(t)$, we can then investigate the time behavior of quantumness and memory of our open qubit under classical control.

\section{Leggett-Garg Inequality}\label{secDM}

In this section, we discuss how the classical driving field can affect the non-classical dynamics of the quantum system. To this aim, we employ Leggett-Garg inequalities which assess the temporal non-classicality of a single system using  correlations measured at different times. 
We recall that the original motivation behind these inequalities was the possibility to check the presence of quantum coherence in macroscopic systems. A Leggett-Garg test entails either the absence of a realistic description of the system or the impossibility of measuring the system without disturbing it. Such inequalities are violated by quantum mechanics on both sides. Some recent theoretical proposals have been reported for the application of Leggett-Garg inequalities in quantum transport, quantum biology and nano-mechanical systems \cite{LG6}.

The simplest Leggett-Garg inequality can be constructed as follows. Consider an observable $\hat{O}$, which can take the values $\pm1$ . One then performs three sets of experimental runs such that in the first set of runs $\hat{O}$ is measured at times $t_{1}$ and $t_{2}=t_{1}+\tau$; in the second, at $t_{1}$ and $t_{3}=t_{1}+2\tau$; in the third, at $t_{2}$ and $t_{3}$. It is noteworthy that such measurements directly enable us to determine the two-time correlation functions $\langle\hat{O}(t_{j})\hat{O}(t_{i})\rangle$ with $t_{j}>t_{i}$. On the basis of the classical assumptions of macroscopic realism and noninvasive measurability, one can derive the Leggett-Garg inequality as \cite{LG1,LG6}
\begin{equation} 
\label{eq:14}  
 C_{3}=C_{21}+C_{32}-C_{31}\leq 1,
\end{equation}                                                                                                  
where $C_{ij}=\langle\lbrace \hat{O}(t_{i}),\hat{O}(t_{j})\rbrace \rangle/2 $, with $\lbrace \hat{O}(t_{i}),\hat{O}(t_{j})\rbrace$ indicating the anticommutator of the two operators.
Utilizing similar arguments, one can also obtain a Leggett-Garg inequality for measurements at four different times, $t_{1}$, $t_{2}$, $t_{3}$ and $t_{4}=t_{1}+3\tau$, given by \cite{LG7,LG8}
\begin{equation}
\label{eq:15}
C_{4}=C_{21}+C_{32}+C_{43}-C_{41}\leq 2.
\end{equation} 
                                                                                          
As mentioned above, the inequalities of Eqs.~(\ref{eq:14}) and (\ref{eq:15}) are obtained under two classical assumptions, so quantum theory violates Leggett-Garg inequalities. Here, we investigate the dynamics of the Leggett-Garg inequalities for our system by choosing the measurement operator $\hat{O}=\hat{\sigma}_{x}$. Exploiting the qubit evolved density matrix of Eq.~(\ref{eq:13}), we obtain
\begin{widetext}
\setlength\arraycolsep{1.4pt}\begin{eqnarray}
\label{eq:16} 
C_{ij}&=&\langle \hat{\sigma}_{x}(t_{i}) \hat{\sigma}_{x}(t_{j})+ \hat{\sigma}_{x}(t_{j}) \hat{\sigma}_{x}(t_{i})\rangle /2 \nonumber \\
&=&\frac{1}{2}( \langle \hat{\sigma}_{+}(t_{j})\hat{\sigma}_{-}(t_{i})+
\hat{\sigma}_{-}(t_{j}) \hat{\sigma}_{+}(t_{i})\rangle  
+\langle \hat{\sigma}_{+}(t_{i}) \hat{\sigma}_{-}(t_{j})+ \hat{\sigma}_{-}(t_{i}) \hat{\sigma}_{+}(t_{j}) \rangle )\nonumber \\
&=& \Re[\cos^{2}(\theta) A(t_{i})A^{*}(t_{j})e^{-i\omega_{D}(t_{i}-t_{j})} 
+ \sin^{2}(\theta)A(t_{i}-t_{j})e^{-i\omega_{D}(t_{i}-t_{j})}],
\end{eqnarray}
\end{widetext}
where $\Re[z]$ indicates the real part of the complex number $z$.

\begin{figure}[b!]
   \centering
\includegraphics[width=0.48\textwidth]{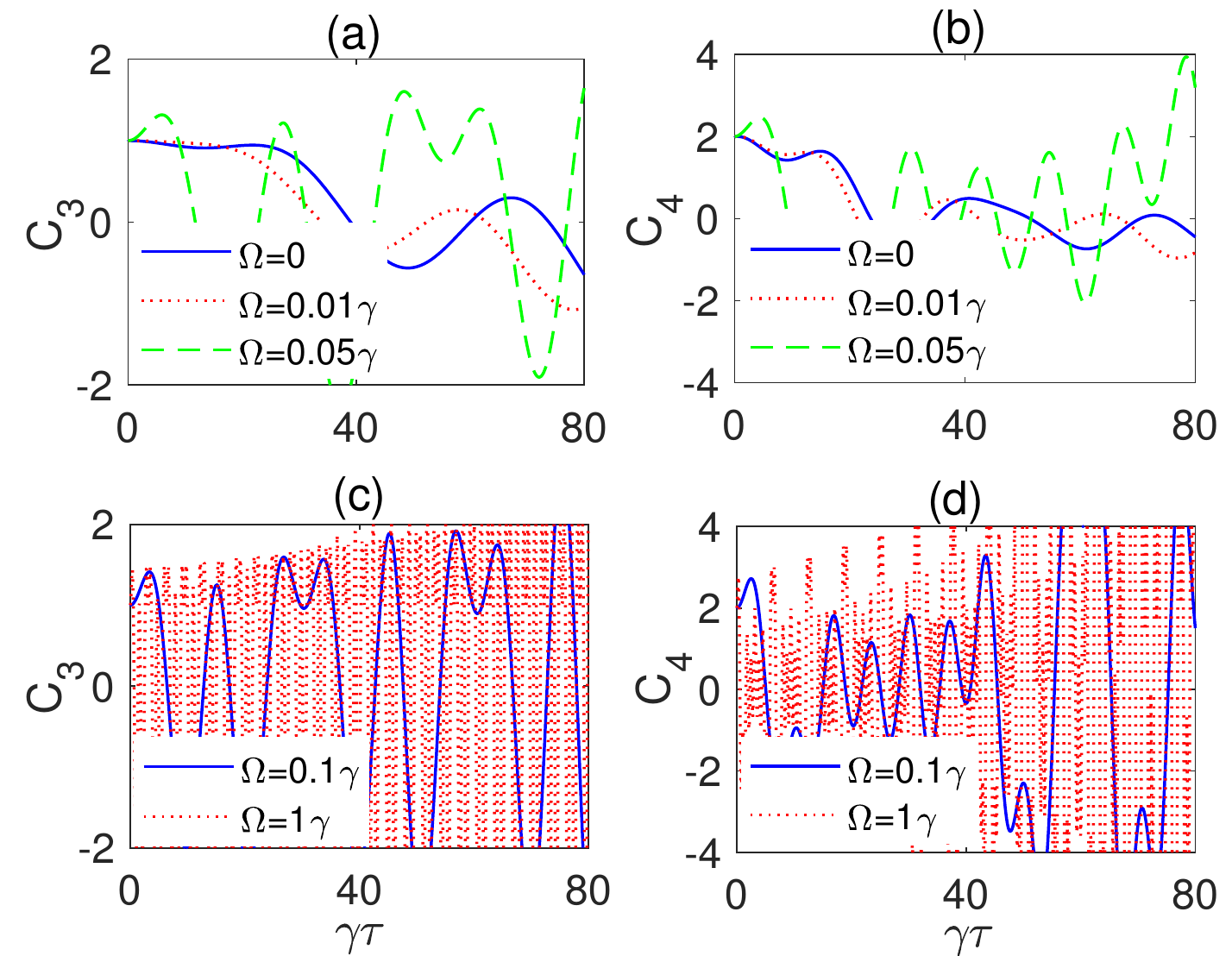}
   \caption{\label{AR2}Leggett-Garg inequalities $C_{3}$ (panels (a) and (c)) and $C_{4}$ (panels (b) and (d)) as functions of the dimensionless time interval $\gamma\tau$ for different values of the qubit-classical field coupling strength (Rabi frequency) $\Omega$. The values of the other parameters are taken as: $\lambda=0.01\gamma$, $\theta=0$, $\delta=\Delta=0$.}
 \end{figure}
                                  
Based on Eq.~(\ref{eq:16}), we calculate and plot in Fig.~\ref{AR2} the dynamics of both Leggett-Garg inequalities $C_{3}$ (panels (a), (c)) and $C_{4}$ (panels (b), (d)), for some values of the coupling strength (Rabi frequency) $\Omega$ between the qubit and the classical field in the resonant case ($\Delta =0$). We also choose a value of the spectral width $\lambda$ such that memory effects are effective ($\lambda=0.01$, strong qubit-cavity coupling) when the classical field is turned off. The qubit initial state is taken to be in the dressed state $\ket{A}$, that is $\theta=0$ in Eq.~(\ref{eq:5}). As can be observed from these plots, in the absence of classical driving field ($\Omega =0$), nonclassical behavior of the qubit (that is, violation of the inequalities) is practically negligible. This means that, despite the presence of memory effects in the dynamics, the manifestation of an experimentally detectable quantum trait of the system by Leggett-Garg inequality violation does not occur without an external control. Contrarily, when the classical driving field is turned on and its intensity increased, we find a significant enhancement of nonclassical behavior ($C_{3}>1$, $C_4>2$) in wide regions of $\gamma\tau$.

\begin{figure}[b!]
   \centering
\includegraphics[width=0.48\textwidth]{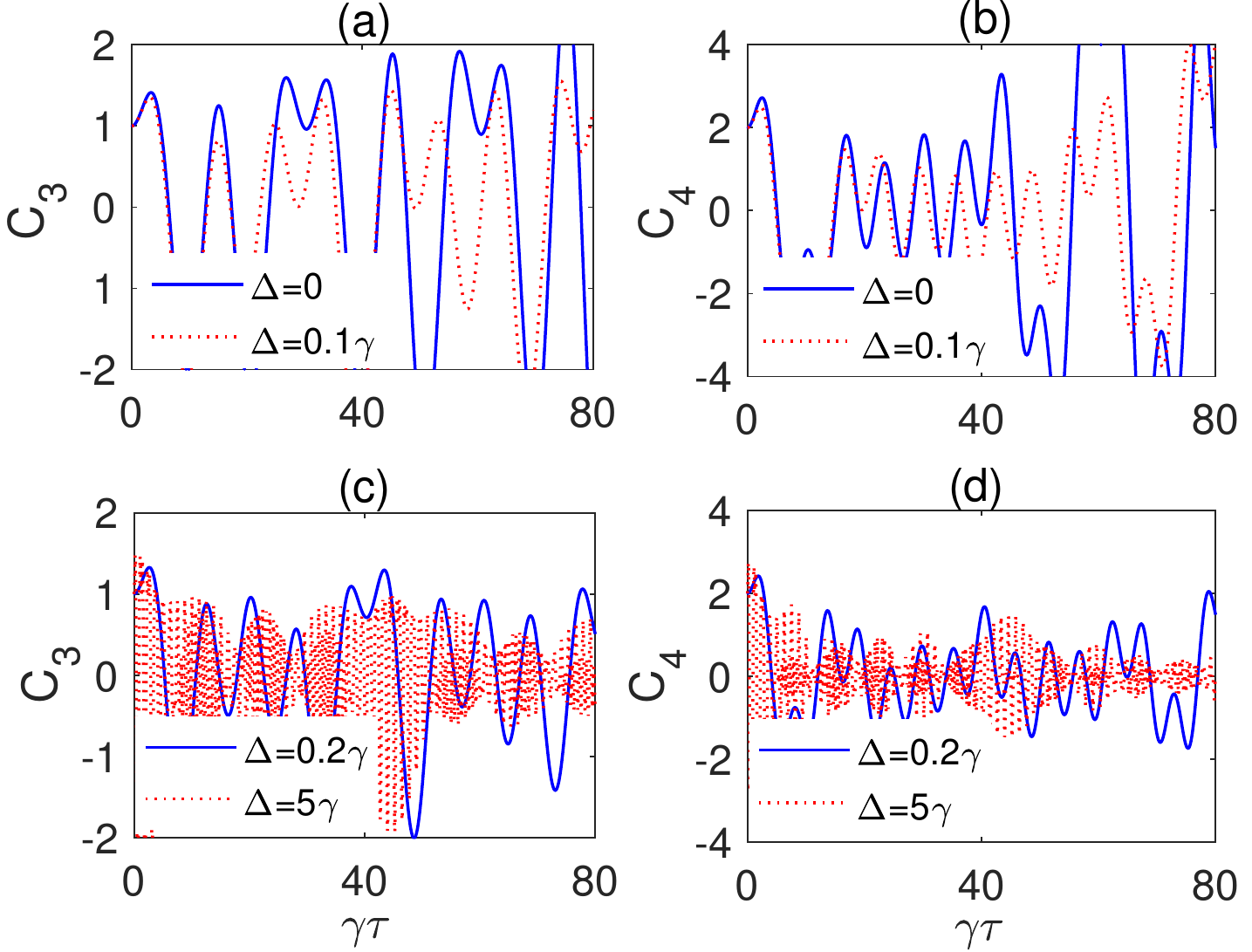}
   \caption{\label{AR3}Leggett-Garg inequalities $C_{3}$ (panels (a) and (c)) and $C_{4}$ (panels (b) and (d)) as functions of the dimensionless time interval $\gamma\tau$ for different values of the qubit-classical field detuning $\Delta$. The values of the other parameters are chosen as: $\lambda=0.01\gamma$, $\theta=0$, $\delta=0$, $\Omega=0.1\gamma$.}
 \end{figure}

To get insights about the role of the qubit-field detuning $\Delta$, we plot in Fig.~\ref{AR3} the time behavior of the Leggett-Garg inequalities $C_{3}$ (panels (a) and (c)) and $C_{4}$ (panels (b) and (d)) for different values of the detuning parameter. The classical control is acting on the system ($\Omega = 0.1\gamma$). It is interesting to notice that the detuning play a detrimental role wit respect to the quantumness of the qubit. In fact, an increase of the detuning significantly reduces the values of the figures of merit for the Leggett-Garg inequalities, which tend to stay below the quantum threshold. As a consequence, the maximum nonclassical behavior of an open qubit is obtained when the classical field resonantly interacts with the quantum system.

\section{Quantum witness and coherence}\label{secQW}

In this section, the quantum features of the controlled qubit are studied using a suitable quantum witness.
Quantum witnesses have been introduced in the literature \cite{LG9, LG11,QWNori} to conveniently probe time-dependent quantum coherence, without resorting to demanding non-invasive measurements or tomographic processes. Such witnesses are particularly effective because they are capable to detect quantumness in a range wider than the Leggett-Garg inequality. These practical advantages make quantum witnesses of experimental interest, with uses not only in open quantum systems but also in quantum transport in solid-state nanostructures \cite{QWNori}. 

Here we adopt the quantum witness defined as \cite{QWNori}
\begin{equation} \label{eq:24}  
W_\mathrm{q}=|p_m(t)-p'_m(t)|,
\end{equation}
where $p_m(t)$ represents the probability of finding the qubit in a state $m$ at time $t$, while $p'_m(t)=\sum^{2}_\mathrm{n=1}p(m,t|n,t_0)p_\mathrm{n}(t_\mathrm{0})$ is the (classical) probability of finding the system in $m$ after a nonselective measurement of the state $n$ has been performed at time $t_0$ (notice that, for a qubit, $d=2$ in the sum). Following the classical no-signaling in the time domain \cite{koflerPRA}, the first measurement at time $t_0$ should not disturb the statistical outcome of the later measurement at time $t$, so that $p_m(t)=p'_m(t)$ ($W_\mathrm{q}=0$ and the system behaves as a classical one. Therefore, a quantum witness $W_\mathrm{q} > 0$ identifies the nonclassicality of the system state at time $t_0$. The conditional probability $p(m,t|n,t_0)$ is expressed as a propagator $\Lambda_\mathrm{mn}(t,t_\mathrm{0})=p(m,t|n,t_0)$, which can be explicitly found by choosing the qubit basis and the state to be measured \cite{LG9}. 

For the controlled open qubit here considered, the natural computational basis is the dressed state basis $\{|A \rangle, \ket{B}\}$ given in Eq.~(\ref{eq:3}) and the qubit is initially ($t=0$) prepared in the coherent superposition $\cos\theta |A\rangle+\sin\theta\ |B\rangle $, as seen in Eq.~(\ref{eq:5}). For calculating the quantum witness, the nonselective measurement of the state $n$ at time $t_0=\tau/2$ is done in the basis $\ket{\pm}=(\ket{A}\pm\ket{B})/\sqrt{2}$, while the state $m$ to be measured at time $t=\tau$ is chosen to be $\ket{+}$ (the projector $\Pi_+=\ket{+}\bra{+}$ is thus measured at $t=\tau$). Exploiting the qubit evolved density matrix of Eq.~(\ref{eq:13}) and determining the explicit expression of the propagator $\Lambda(t,t_\mathrm{0})=||\Lambda_\mathrm{mn}(t,t_\mathrm{0})||$, the quantum witness of Eq.~(\ref{eq:24}) is (see Appendix~\ref{App})
\begin{equation} \label{eq:Wq}  
W_\mathrm{q}=\frac{1}{4}\left|\sin(2\theta)\left(A(\tau)+A^{*}(\tau) - \frac{1}{2}(A(\tau/2)+A^{*}(\tau/2))^2 \right)\right|,
\end{equation}
where $A(t)$ is given in Eq.~(\ref{eq:11}).

\begin{figure}[t!]
 \centering
\includegraphics[width=0.48\textwidth]{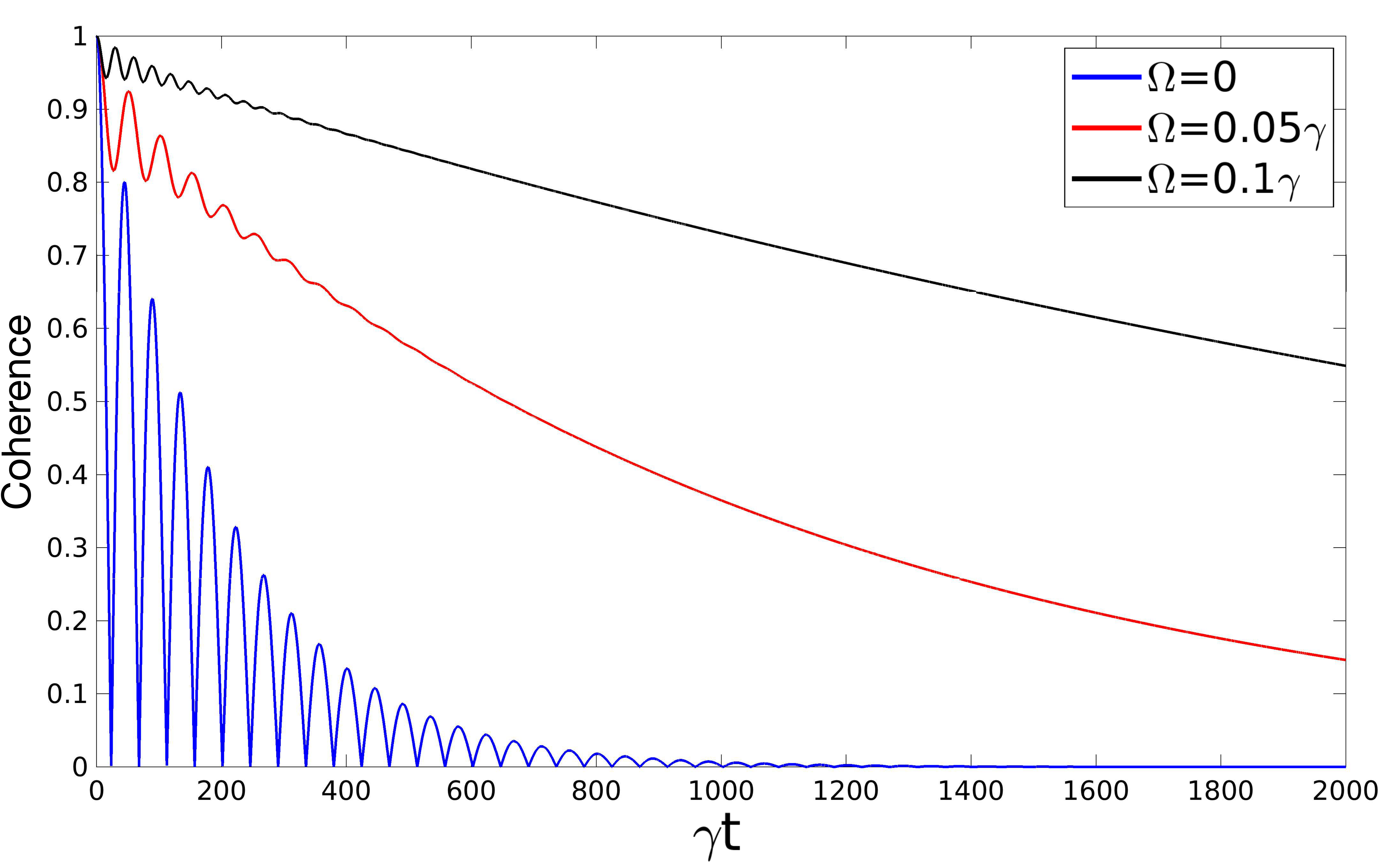}
\caption{Coherence $C_{l_1}(t)$ as a function of the dimensionless time  
$\gamma t$ for different values of $\Omega$, with $\lambda=0.01\gamma$ and $\Delta=0$. Other parameters are $\theta=\pi/4$ (qubit initial state $\ket{+}$) and $\delta=0$.}
\label{fig:coh}
\end{figure}

\begin{figure*}[t!]
 \centering
\includegraphics[width=0.9\textwidth]{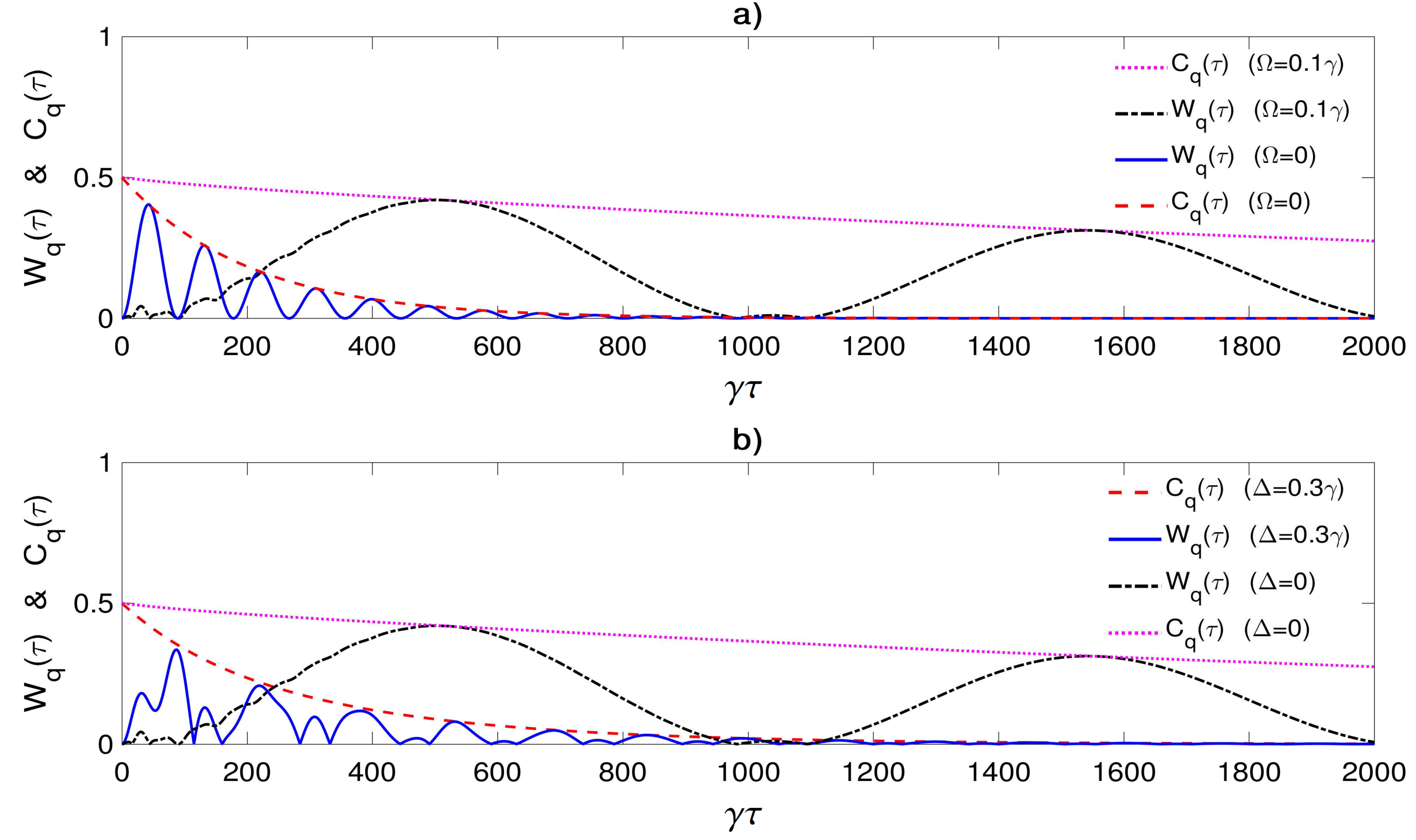}
\caption{Quantum witness $W_{q}(\tau)$ and coherence monotone $C_\mathrm{q}(\tau)$ (envelope of half quantum coherence) as a function of the dimensionless time interval $\gamma\tau$ for (a) different values of $\Omega$ with $\Delta=0$, and (b) different values of detuning $\Delta$ with $\Omega=0.1\gamma$. Other parameters: $\lambda=0.01\gamma$, $\theta=\pi/4$ (qubit initial state $\ket{+}$) and $\delta=0$.}
\label{fig:QWCoh}
\end{figure*} 

In general, as mentioned above, the quantum witness provides qualitative information whether the qubit exhibits quantum behavior during the evolution. In view of a more quantitative inspection, giving us the amount of quantumness the qubit retains during the dynamics, a quantifier of quantum coherence is needed \cite{streltsovRMP}. By using the well-known $l_1$-norm of coherence $C_{l_1}(t)=\sum_{i\neq j} |\rho_{ij}|$ \cite{baumgratzPRL}, from the qubit reduced density matrix of Eq. \eqref{eq:13} one easily obtains $C_{l_1}(t)=|A(t)|$. It is also interesting to compare in our system the dynamics of $W_\mathrm{q}(\tau)$ with the so-called coherence monotone $C_\mathrm{q}(\tau)$, defined as the envelope of $C_{l_1}(\tau)/2$, which has been shown to provide the upper bound of the quantum witness in a spontaneous Markovian (memoryless) decay of a qubit in a thermal reservoir \cite{LG9}. In Figures \ref{fig:coh} and \ref{fig:QWCoh} we plot, respectively, the dynamics of quantum coherence $C_{l_1}(t)$ and the dynamics of $W_\mathrm{q}(\tau)$ together with the coherence monotone, starting from the qubit (coherent) initial state $\ket{+}$ ($\theta=\pi/4$ in Eq. \eqref{eq:5}) under qubit-cavity strong-coupling ($\lambda=0.01\gamma$) and different values of other relevant parameters. As expected, coherence is better protected by increasing the coupling to the control of classical driving field (resonant case $\Delta =0$).
From Fig.~\ref{fig:QWCoh}(a-b), we find that in this model the maximum of the quantum witness coincides with the coherence monotone, showing a quantitative relation between witness ($W_\mathrm{q}$) and quantifier ($C_{l_1}$) of quantumness. The quantum witness decays rapidly in the absence of the classical driving field ($\Omega=0$), although there are significant oscillations above zero which allows us to detect quantumness anyway. This shows the finer resolution of $W_\mathrm{q}$ as an indicator of nonclassicality with respect to the Leggett-Garg inequality under the same condition. 
However, qubit quantumness can be maintained for longer times by increasing the coupling $\Omega$ to the resonant ($\Delta=0$) classical control (see Fig.~\ref{fig:QWCoh}(a)), a behavior similar to that of the Leggett-Garg inequality violation shown in Sec.~\ref{secDM}. Therefore, $W_\mathrm{q}$ quantitatively confirms that the controllable classical field leads to extending the lifetime of quantum coherence under resonant interaction. 
It is then natural to explore the role of the detuning $\Delta$. This is displayed in Fig.~\ref{fig:QWCoh}(b), fixing a qubit-cavity strong coupling $\lambda=0.01\gamma$ and an interaction strength $\Omega=0.1\gamma$ with the classical control (corresponding to the best performance in the resonant case of Fig.~\ref{fig:QWCoh}(a)). As a general behavior, we find that the maintenance property of quantum witness (and therefore coherence) is significantly weakened by increasing $\Delta$. An out-of-resonance classical control is thus adverse to a long-time manifestation of quantumness for the qubit. 

\begin{figure}[t!]
   \centering
{\includegraphics[width=0.48\textwidth]{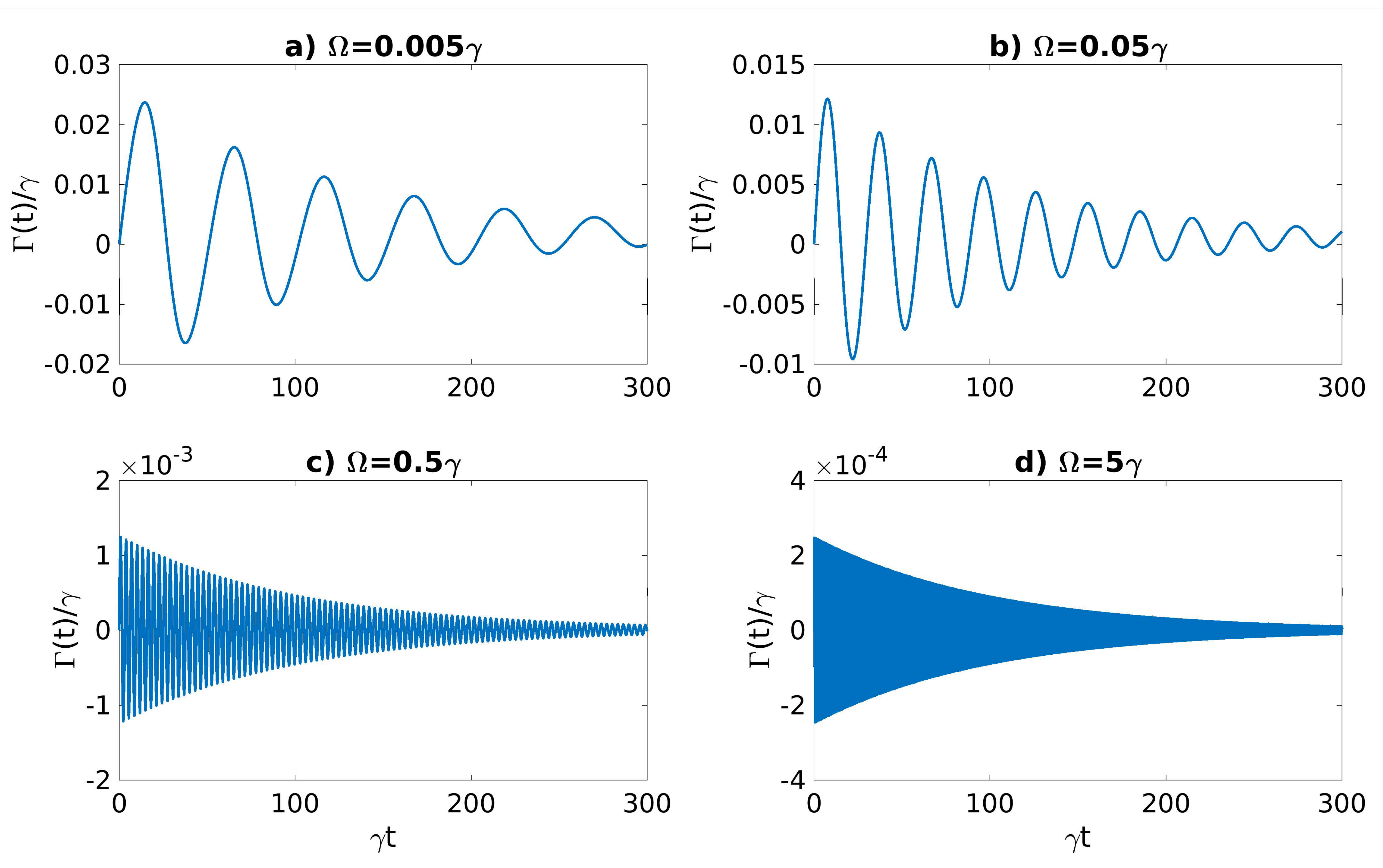}}
   \caption{Effective decay rate $\Gamma(t)$ (in units of $\gamma$) as a function of the dimensionless time $\gamma t$ for different values of Rabi frequency of classing driving field $\Omega$, with $\lambda=0.01\gamma$ and $\Delta=0$. Other parameters are $\theta=\pi/4$ (qubit initial state $\ket{+}$) and $\delta=0$.}
\label{fig:Gammat}
 \end{figure}

The above results can be interpreted by an analogy to dynamical decoupling technique \cite{viola1999DD,DDreview}. Indeed, the classical field acting on the open system (qubit) can be viewed as responsible of a controllable cyclic decoupling interaction term $H_\mathrm{D}:=\Omega e^{-i\omega_{c}t}\hat{\chi}_{+}+c.c.$ in the Hamiltonian of Eq. \eqref{eq:1}, which can shield the qubit from noise. This analogy can be confirmed by looking at the behavior of the effective time-dependent decay rate arising from the qubit reduced density matrix of Eq. \eqref{eq:13}, that is \cite{NM1,NM13} $\Gamma(t)=-2\mathrm{Re}\{\dot{A}(t)/A(t)\}$. Figure \ref{fig:Gammat} displays that, under resonant interaction, increasing the coupling $\Omega$ of the qubit to the external driving field enables a reduction of $\Gamma(t)$, implying an inhibition of the detrimental effects of noise.

\section{Geometric phase}\label{secU}

In this section, we analyze the influence of the classical driving field on the geometric phase (GP) of the system. We already mentioned in the Introduction how the GP and its properties have found applications in the context of geometric quantum computation, allowing quantum gates with highest fidelities \cite{GP28,GP29,GP30,GP31,GP32}. One of the principal motivations for having quantum gates based on GP is that they seem to be more robust against noise and errors than traditional dynamical gates, thanks to their topological properties \cite{GPQI,pachosPRL,pachosPRA,mollerPRA}. Geometric quantum computation exploits GPs to implement universal sets of one-qubit and two-qubit gates, whose realization finds versatile platforms in systems of trapped atoms (or ions) \cite{Duan1695,recati2002}, quantum dots \cite{solinasPRB} and superconducting circuit-QED \cite{GP30,GP31,GP32,faoroPRL,filippNature}. Since our system can be implemented in these experimental contexts, using real or artificial atoms, it is important to unveil the time behavior of the qubit geometric phase under classical control.

To this aim, we need to determine the geometric phase for a mixed (noisy) state of a qubit \cite{GPQI}. In particular, we adopt the kinematic method \cite{GP9} to calculate the geometric phase of the qubit which undergoes a non-unitary evolution. According to this method, the qubit GP is given by \cite{GP9}
\begin{equation}
\label{eq:17}
\Phi_{g}=\arg\left\lbrace \sum_{i}\sqrt{\varepsilon_{i}(0)\varepsilon_{i}(T)}\langle\varepsilon_{i}(0)|\varepsilon_{i}(T)\rangle e^{\int_{0}^{T}dt\langle\delta_{i}(t)|\frac{\partial}{\partial t}| \delta_{i}(t)\rangle}\right\rbrace , 
\end{equation}
where $\varepsilon_{i}(0)$, $\varepsilon_{i}(T)$ and $| \varepsilon_{i}(0)\rangle$, $| \varepsilon_{i}(T)\rangle$ ($i=1,2$) are the instantaneous eigenvalues and eigenvectors of the reduced density matrix of the qubit at times $t=0$, $T$, respectively. 

In order to get the desired qubit geometric phase for our system, we first calculate the eigenvalues and eigenstates of the density matrix $\rho(t)$ of Eq.~(\ref{eq:13}) which are, respectively,
\begin{equation} 
\label{eq:18}  
\varepsilon_{\pm}(t)=\frac{1}{2}\left\lbrace  1\pm
\sqrt{\vert A(t)\vert^{2}\sin^{2}(2\theta)+
(2\vert A(t)\vert^{2}\cos^{2}\theta - 1 )^{2}}\right\rbrace,
\end{equation}                                             
and                                             
\setlength\arraycolsep{1.4pt}\begin{eqnarray}
\label{eq:19} 
| \varepsilon_{+}(t)\rangle&=& e^{-i\omega_{D}t/2}\cos\Theta | E\rangle +e^{-i\omega_{D}t/2}\sin\Theta | G\rangle ,\nonumber \\
| \varepsilon_{-}(t)\rangle&=& -e^{-i\omega_{D}t/2}\sin\Theta | E\rangle +e^{-i\omega_{D}t/2}\sin\Theta | G\rangle,
\end{eqnarray}                                                                 
where 
\begin{equation} 
\label{eq:20}  
\cos\Theta=\frac{2(\vert A(t)\vert^{2}\cos^{2}\theta - \varepsilon_{-})}{\sqrt{\vert A(t)\vert^{2}\sin^{2}(2\theta)+4(\vert A(t)\vert^{2}\cos^{2}\theta - \varepsilon_{-}. )^{2} }}.
\end{equation}
It is clear that $\varepsilon_{-}(0)=0$. Consequently, in this case only the ``$+$'' mode contributes to the GP, as evinced from Eq.~(\ref{eq:17}). Then, the GP of the qubit after a period $T=2\pi/\omega_{D}$ can be readily obtained by Eq.~(\ref{eq:17}) as
\begin{equation} 
\label{eq:21}  
\Phi_{g}=\omega_{D}\int_{0}^{T}\cos^{2}\Theta dt.
\end{equation}
       
\begin{figure}[t!]
   \centering
\includegraphics[width=0.47\textwidth]{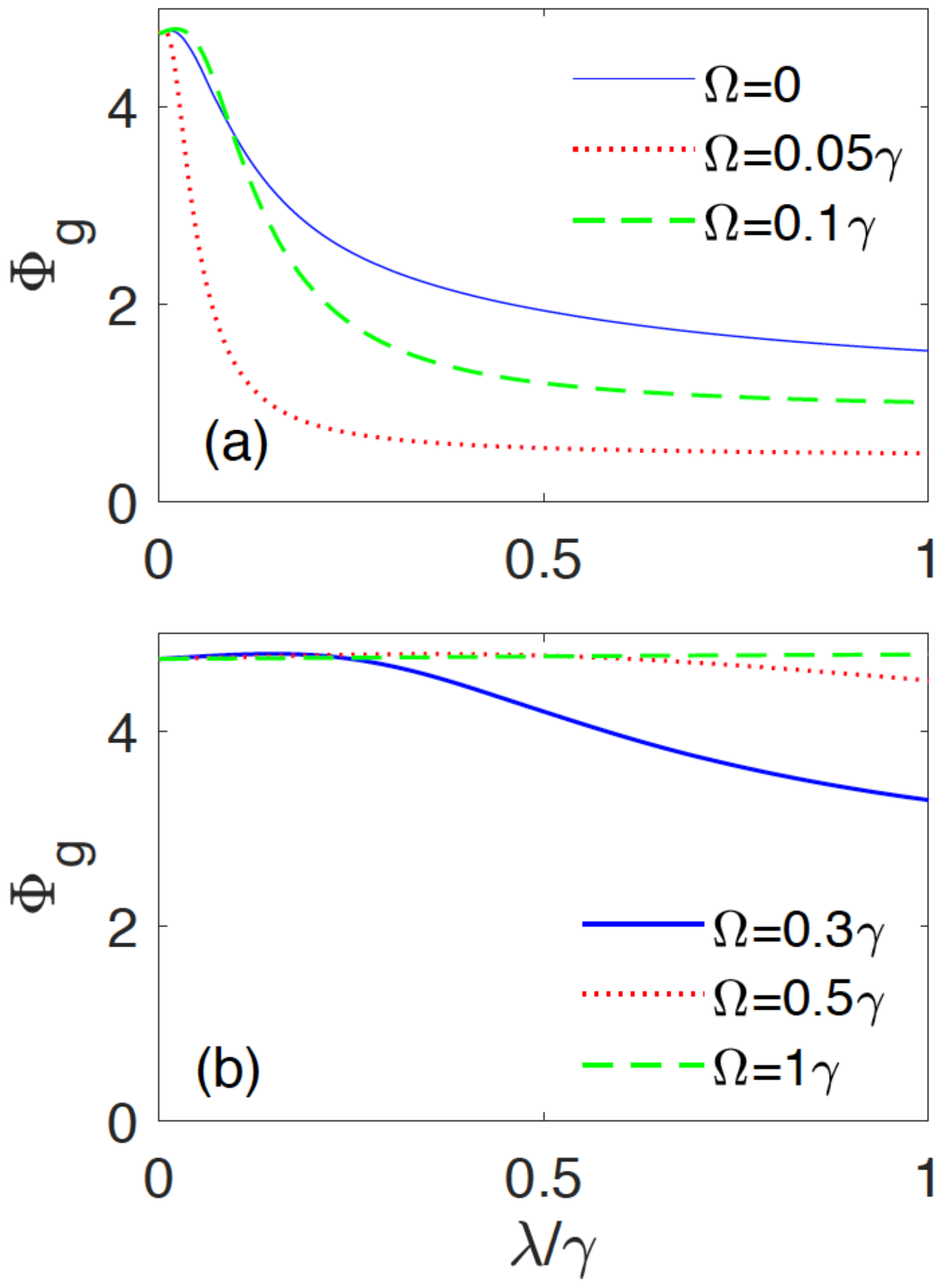}
   \caption{\label{AR4}Geometric phase $\Phi_{g}$ of the qubit, after a period $T=2\pi/\omega_{D}$, as a function of the (scaled) cavity spectral width $\lambda/\gamma$ for different values of the qubit-classical field coupling strength (Rabi frequency) $\Omega$. Other parameters are taken as $\theta=\pi /6 $, $\delta=\Delta=0$.}
 \end{figure}
                                                                                         
Fig.~\ref{AR4} displays the behavior of GP versus $\lambda/\gamma$ for different coupling strengths $\Omega$ of the classical driving field with the qubit, under resonant interaction. The qubit initial state is taken as a superposition of the two dressed states, choosing 
$\theta=\pi /6$ in Eq.~(\ref{eq:5}). Such a study allows us to show the interplay between the couplings of the qubit with both cavity field ($\lambda$) and classical field ($\Omega$). In general, whatever the value of $\Omega$, the geometric phase tends to diminish with the increase of $\lambda/\gamma$, eventually approaching an almost fixed value. So, increasing the cavity spectral width 
$\lambda$ (that is, decreasing the cavity quality factor and the qubit-cavity coupling) reduces the value of the geometric phase acquired from the qubit. A remarkable aspect is provided, for an assigned value of $\lambda/\gamma$, by the non-monotonic behavior of the qubit GP when $\Omega$ is increased. As seen from Fig.~\ref{AR4}(a-b), the GP firstly becomes smaller when the classical control is turned on ($\Omega\leq 0.1\gamma$, panel (a)), and then starts increasing for larger intensities of the coupling ($\Omega\geq 0.3\gamma$, panel (b)), significantly overcoming the value reached in absence of the classical field. We point out how the geometric phase of the qubit evolved mixed state can be thus efficiently stabilized, independently of the cavity spectral width, by suitably adjusting the coupling of the qubit to the classical control (see, for instance, $\Omega=1\gamma$). This feature is quite useful, since it entails that there is no need to have a high-$Q$ cavity to maintain a given amount of geometric phase, provided that an intense resonant classical control field is tailoring the qubit.

\begin{figure}[t!]
   \centering
\includegraphics[width=0.47\textwidth]{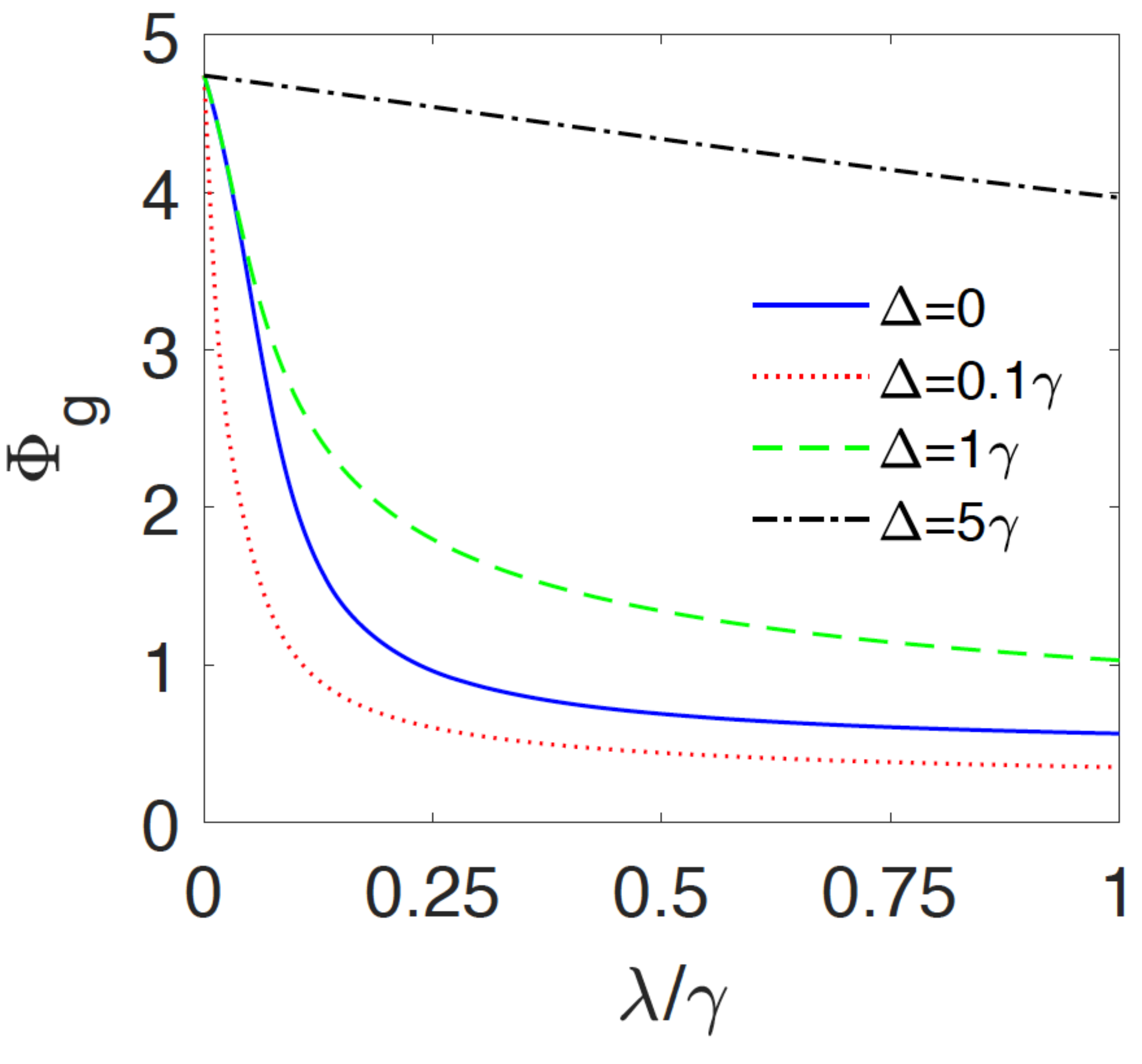}
   \caption{\label{AR5}Geometric Phase ($\Phi_{g}$) as a function of spectral width $\lambda/\gamma$ for different values of detuning frequency of classing driving field ($\Delta$). Other parameters are $\theta=\pi /6$, $\delta=0$, $\Omega=0.1\gamma$.}
 \end{figure}

The effects due to an out-of-resonance classical control are then shown in Fig.~\ref{AR5}. The curves here display the GP versus $\lambda/\gamma$ for different values of the detuning parameter $\Delta$, for a small coupling to the classical field ($\Omega=0.1\gamma$). We choose a small value of $\Omega=0.1\gamma$ to understand whether the detuning is capable to enrich or not the value of the GP. Also in this case we retrieve a non-monotonic behavior with respect to the detuning, for a fixed value of $\lambda/\gamma$. In fact, it is evident from Fig.~\ref{AR5} that when the classical field is near resonant ($0<\Delta\leq 0.1\gamma$) the acquired GP becomes more fragile than the resonant case ($\Delta=0$), while for the case far from resonance the GP decreases more slowly with respect to the growth of the cavity spectral width $\lambda/\gamma$. Therefore, for a small qubit-classical field coupling, a non-resonant control is more convenient to stabilize the geometric phase of the open qubit to a larger amount.

\section{Non-Markovianity}\label{secNon}

In this section we focus on the memory effects present in the system dynamics, establishing their dependence on the classical field. We recall that without the external control ($\Omega=0$), memory effects due to non-Markovianity are physically characterized by the ratio $\lambda/\gamma$. In fact, strong coupling conditions for which $\lambda/\gamma<1$ activate a non-Markovian regime for the system dynamics due to a cavity correlation time larger than the qubit relaxation time \cite{NM1,lofrancoreview}. On the contrary, values of $\lambda/\gamma>1$ usually identify a weak coupling with Markovian dynamics. However, the action of the external classical field may modify the conditions for which memory effects are present, which therefore deserve to be investigated.

To identify non-Markovian dynamics of the system, among the various quantifiers, we utilize the so-called $BLP$ measure \cite{NM4} that is based on the distinguishability between two evolved quantum states, measured by the trace distance. We briefly recall the main physical ingredients behind this non-Markovianity measure. For any two states $\rho_{1}(t)$ and $\rho_{2}(t)$ undergoing the same evolution, their trace distance is
\begin{equation} \label{eq:22}  
D[\rho_{1}(t), \rho_{2}(t) ]=( 1/2)\mathrm{Tr}\vert\rho_{1}(t)-\rho_{2}(t)\vert,
\end{equation}                                                                                
where $\vert X \vert=\sqrt{X^{\dagger}X}$ and $0\leq D\leq 1$ \cite{LG4}. 
The time derivative of the trace distance ($\sigma[t, \rho_{1}(0), \rho_{2}(0) ]=\mathrm{d}D [\rho_{1}(t), \rho_{2}(t) ]/\mathrm{d}t$) can be interpreted as characterizing a flow of information between the system and its environment.
Within this scenario, Markovian processes satisfy $\sigma\leq 0$ for all pairs of initial states $\rho_{1,2}(0)$ at any time $t$. Namely, $\rho_{1}(t)$ and $\rho_{2}(t)$ will eventually lose all their initial information and become identical. However, when $\sigma>0$, $\rho_{1}(t)$ and $\rho_{2}(t)$ are away from each other. This can be seen as a backflow of quantum information from the environment to the system and the process is called non-Markovian.
Based on this concept, a measure of non-Markovianity can be defined by \cite{NM4}
\begin{equation} \label{eq:23}  
\mathcal{N}=\mathrm{max}_{\rho_{1}(0), \rho_{2}(0)}\int_{\sigma>0}\sigma[t, \rho_{1}(0), \rho_{2}(0)]\mathrm{d}t,
\end{equation}
where the time integration is taken over all intervals in which there is information backflow ($\sigma>0$) and the maximization is done over all possible pairs of initial states.

\begin{figure}[t!]
   \centering
\includegraphics[width=0.48\textwidth]{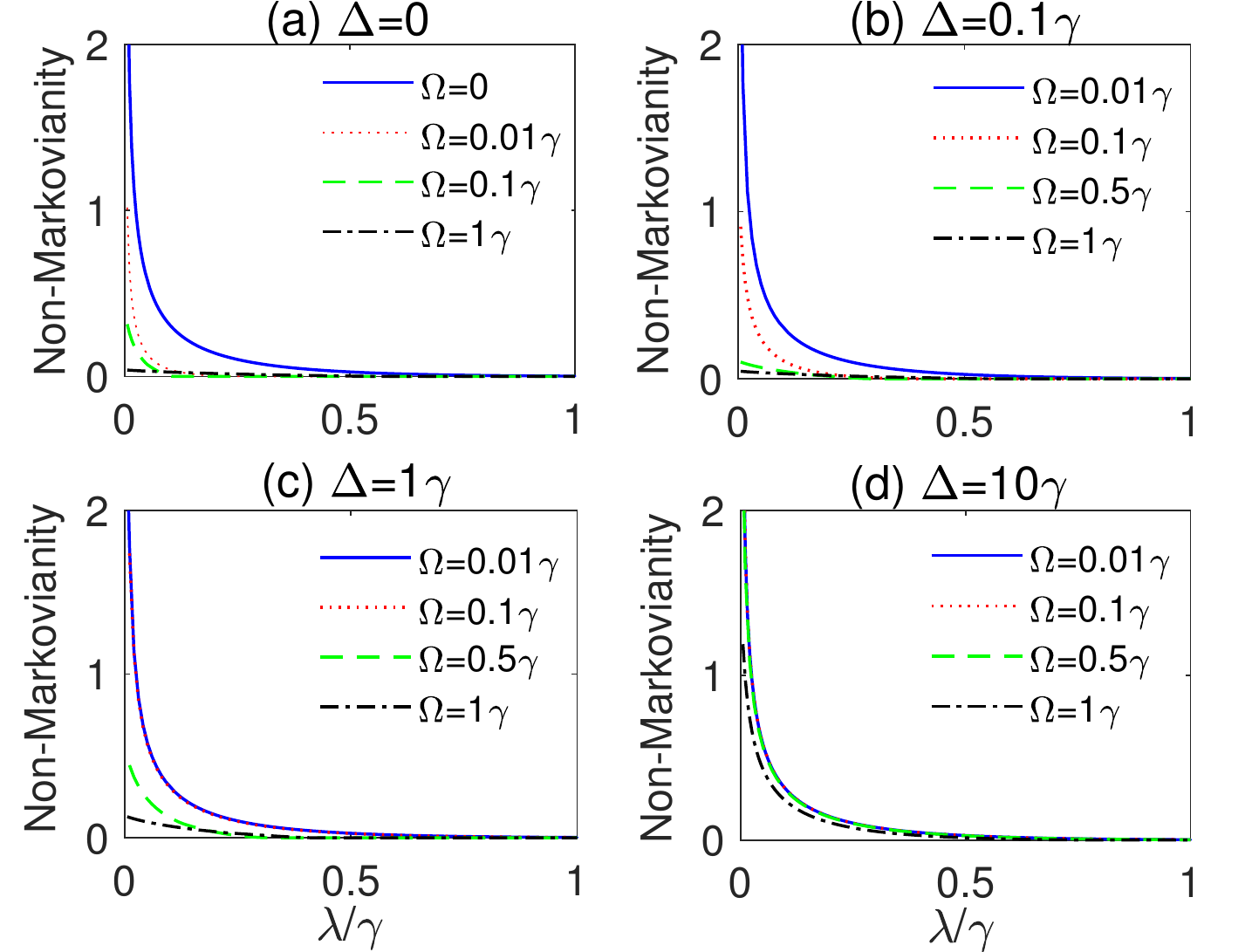}
   \caption{\label{AR6}Non-Markovianity as a function of $\lambda/\gamma$ for different $\Omega$; (a) $\Delta=0$, (b) $\Delta=0.1\gamma$, (c) $\Delta=1\gamma$ and (d) $\Delta=10\gamma$. The values of the other parameters are taken as: $\theta=0$,  $\delta=0$.}
 \end{figure}  
                                                                                                        
In Fig.~\ref{AR6}, we display the effect of the coupling strength of classical driving field to the qubit on the non-Markovianity as a function on the ratio $\lambda/\gamma$ for different detuning $\Delta$. The initial state of the qubit is chosen to be in the dressed state $\ket{A}$, corresponding to $\theta=0$ in Eq.~(\ref{eq:5}). In the absence of classical field ($\Omega=0$) the known behavior explained above is quantitatively retrieved, that is memory effects significantly decrease with an increase of $\lambda/\gamma$ (a larger cavity spectral width). The plots of Fig.~\ref{AR6} then clearly show that the action of the classical control tends to destroy non-Markovianity, especially in the resonant case $\Delta=0$ (see panel (a)). An out-of-resonance interaction mitigates the disappearance of memory effects, whose amount of non-Markovianity approaches that occurring without classical field (see panel (d), in particular). We highlight that such a behavior is in contrast with the enriching effect of the classical control field on the quantum features of the qubit, such as Leggett-Garg inequality, quantum witness and geometric phase treated above. This aspect makes it emerge that larger memory effects do not correspond to enhanced quantumness of the system, contrarily to what one may expect. In fact, weaker non-Markovianity here occurs in correspondence to enriched dynamics of quantumness of the controlled qubit.   

\begin{figure}[t!]
   \centering
\includegraphics[width=0.5\textwidth]{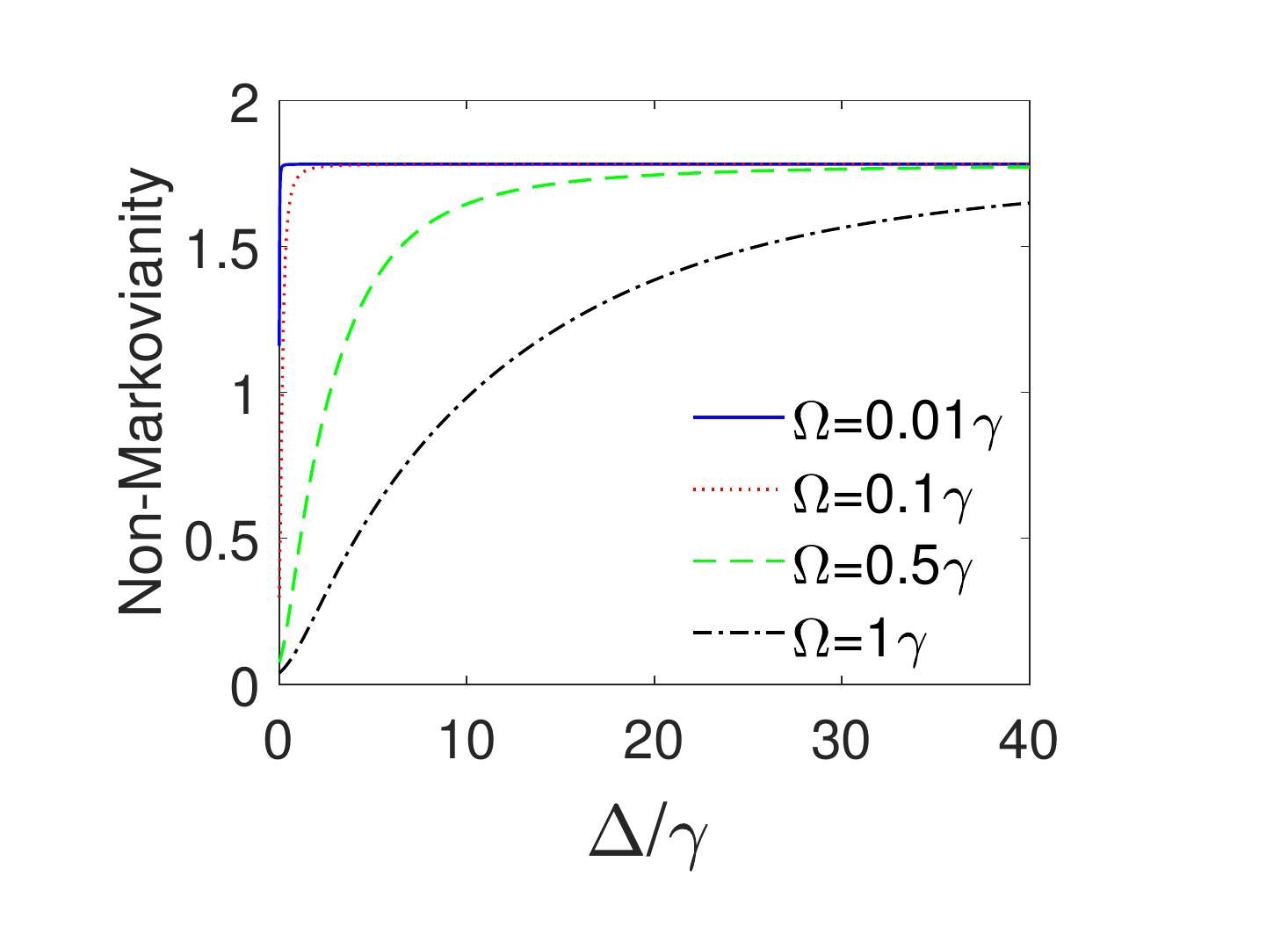}
   \caption{\label{AR7}Non-Markovianity as a function of $\Delta/\gamma$  for different $\Omega$; $\Omega=0.01\gamma$  (solid blue line), $\Omega=0.1\gamma$ (dotted red line), $\Omega=0.5\gamma$ (dashed green line), $\Omega=1\gamma$ (dash-dotted black line). The values of the other parameters are taken as: $\lambda=0.01\gamma$, $\delta=0$, $\theta=0$.}
 \end{figure} 

To better figure out the interactive role of the coupling constant (Rabi frequency) $\Omega$ and detuning $\Delta$ of the classical driving field, we also plot non-Markovianity as a function of $\Delta/\gamma$ for different values of $\Omega$ in Fig.~\ref{AR7}. The curves here confirm that an intense classical field strongly reduces non-Markovianity of the system, especially in the resonant case. To prevent this, large detuning is required. Namely, the larger $\Omega$, the higher the values of the detuning required in order to maintain a given degree of non-Markovianity.

\section{Conclusion}\label{secCon}

In this work, we have investigated the role of an external classical field as a control of quantumness, geometric phase and non-Markovianity of a qubit embedded in a leaky cavity. We focused on the dynamics of fundamental quantum traits linked to the superposition principle (coherence) of the qubit which are experimentally detectable in a direct way, identified by Leggett-Garg inequality violations and a suitable quantum witness. Considering such quantities is advantageous from a practical perspective, since quantumness can be detected without using tomographic techniques, which require experimental resources in terms of measurement settings that increase exponentially with the system complexity \cite{QWNori}. 

A general remarkable point we have found is that an intense and resonant classical control is required to enrich the quantumness of a qubit during its time evolution. The more significant the interaction of the quantum system with an external classical field, the more quantum the system behaves. As a consequence of the quantumness enhancement, we have then shown that the geometric phase acquired by the qubit during a period of its noisy evolution can be efficiently stabilized, independently of the cavity spectral width, by suitably adjusting the coupling of the qubit to the classical control. Cavities with very high quality factors are therefore not needed to maintain a given geometric phase of the open qubit, provided that the latter is harnessed by an intense resonant classical field. We have also seen that the action of the classical control strongly weakens non-Markovianity (memory effects), especially in the resonant case, in contrast with the enriching effect it has onto the quantum behavior of the qubit, such as Leggett-Garg inequality, quantum witness and geometric phase. This aspect makes it clear that larger memory effects do not necessarily entail enhanced quantumness of the system. Therefore, quantumness protection is not due to system-environment information backflow but it is rather caused by the controllable interaction of the open qubit with the external classical field which, as we have shown, induces a reduction of the effective time-dependent decay rate. 

The system considered and the corresponding findings can be feasible by current technologies in both cavity and circuit-QED, which enable the manipulation of interactions of classical fields with qubits \cite{Paik2011PRL,tuorilaPRL,liNatComm,beaudoinPRA}. Our results supply useful insights towards a deeper comprehension of the interaction of a classical signal with a quantum device, highlighting the importance of hybrid classical-quantum systems in a quantum information scenario. 


\appendix

\section{Expression of the state evolution amplitude $B_{k}(t)$}
\label{amplitudeB}

Analogously to the passages which conduct to the determination of $A(t)$, an integro-differential equation can be obtained for the probability amplitude $B_{k}(t)$ of Eq. \eqref{eq:6} \cite{NM1}. Its explicit expression, which does not play any role in our analysis, is
\setlength\arraycolsep{1.4pt}\begin{eqnarray}
\label{eq:12} 
B_{k}(t) &=& -i\frac{g^{*}_{k}\cos\theta }{2}\cos^{2}(\eta/2)\nonumber \\
&&\times\left\{\left(1+\frac{2M}{F} \right) \left(\frac{1-e^{-[\frac{-M}{2}-\frac{F}{4}-i(\delta_{k}+\Delta - \omega_{D})] t}}{\frac{M}{2}-\frac{F}{4}-i(\delta_{k} +\Delta -\omega_{D})}\right) \right. \nonumber \\
&&\left. +\left(1-\frac{2M}{F}\right) \left(\frac{1-e^{-[\frac{M}{2}+\frac{F}{4}-i(\delta_{k} +\Delta -\omega_{D})] t}}{\frac{M}{2}+\frac{F}{4}-i(\delta_{k}+\Delta -\omega_{D})}\right)\right\},
\end{eqnarray}                                
where $\delta_{k}=\omega_{k}-\omega_{0}$ .

\section{Calculation of the quantum witness} \label{App}

In this appendix, we report the evaluation of the quantum witness $W_\mathrm{q}$ of Eq.~(\ref{eq:24}). 

To compute the propagator $\Lambda$, we first need to define the Lindblad-type evolution of an operator $\hat{X}$ within the Heisenberg picture, that is $\mathrm{d}\hat{X}/\mathrm{d}t=\mathcal{L}[\hat{X}]$ \cite{LG9,LG11}. Assuming the cavity owns a single photon in mode $k$, the Lindblad operators can be defined in a basis consisting of raising and lowering Pauli operators ($\hat{\sigma}_{+}=\hat{\sigma}_{x}+i\hat{\sigma}_{y}$ and $\hat{\sigma}_{-}=\hat{\sigma}_{x}-i\hat{\sigma}_{y}$). Notice that the computational basis for our system is the dressed state basis $\{\ket{A},\ket{B}\}$ given in Eq.~(\ref{eq:3}), where $\hat{\sigma}_{-}=| B\rangle \langle A|$ ($\hat{\sigma}_{+}=| A\rangle \langle B|$) represents the lowering (raising) qubit operator. The integro-differential equation for the operator $\hat{X}$, arising from the Lindblad equation, is expressible as $\hat{X}+L[\hat{X}]=0$. For our system, this equation takes the form
\begin{equation}
\hat{X}(t)+ \cos^{4}(\eta/2)\int_{0}^{t}\mathrm{d}t'\mathcal{F}_t[\hat{X}(t')]= 0,  
\end{equation} 
where
\begin{equation}
\mathcal{F}_t[\hat{X}(t')]=F(t,t^{\prime})(\hat{\sigma}_{+}\hat{\sigma}_{-}\hat{X}(t^{\prime})+\hat{X}(t^{\prime})\hat{\sigma}_{+}\hat{\sigma}_{-}-2\hat{\sigma}_{+}\hat{X}(t^{\prime})\hat{\sigma}_{-}),
\end{equation} 
with the function $F(t,t^{\prime})$ being the kernel (correlation function) of 
Eq.~(\ref{eq:8}).

When the operator $\hat{X}$ is substituted by the projectors $\Pi_\pm=\frac{1}{2}(\openone \pm \sigma_x)$ onto the eigenstates $\ket{\pm}=(\ket{A}\pm\ket{B})/\sqrt{2}$ of $\sigma_x$, we obtain 
\begin{eqnarray}\footnotesize
\left(\begin{array}{cc} \Pi_\mathrm{+}(t) \\ \Pi_\mathrm{-}(t) \end{array}\right)+
\cos^{4}(\eta/2)\left(\begin{array}{cc}
\int^t_{0}dt^{\prime}F(t,t^{\prime})     &-\int^t_{0}dt^{\prime}F(t,t^{\prime})   \\
 -\int^t_{0}dt^{\prime}F(t,t^{\prime})     & \int^t_{0}dt^{\prime}F(t,t^{\prime}) 
\end{array}\right)\left(\begin{array}{cc} \Pi_\mathrm{+}(t^{\prime}) \\ \Pi_\mathrm{-}(t^{\prime}) \end{array}\right)=0.\nonumber \\
\end{eqnarray}

From the above equation, the propagator $\Lambda(t,0)$ in the basis $\{\ket{\pm}\}$, such that $\Pi_\pm(t)=\Lambda(t,0)\Pi_\pm(0)$ is finally obtained as 
\begin{equation}
\Lambda(t,0)=\frac{1}{2}\left(\begin{array}{cc} 1+\frac{1}{2}(A(t)+A^{*}(t)) &  1-\frac{1}{2}(A(t)+A^{*}(t))\\ 
1-\frac{1}{2}(A(t)+A^{*}(t)) & 1+\frac{1}{2}(A(t)+A^{*}(t))
\end{array}\right),
\end{equation}
whose matrix elements are $\Lambda_{mn}(t,0)$ with $(m,n)=(+,-)$.

In absence of the intermediate nonselective measurement, the quantum probability $p_\pm$ of finding the state $\ket{\pm}$ at time $\tau$ is defined by $p_\pm(\tau)=\langle\Pi_\pm(\tau)\rangle=\mathrm{Tr}(\rho(\tau)\Pi_\pm(\tau))$, where $\rho(\tau)$ is the evolved reduced density matrix of the qubit. Using the propagator $\Lambda(t,0)$ above, this probability results to be
\begin{equation}
    \left(\begin{array}{cc} p_\mathrm{+}(\tau) \\ p_\mathrm{-}(\tau) \end{array}\right)=\Lambda(\tau,0)\left(\begin{array}{cc} p_\mathrm{+}(0) \\ p_\mathrm{-}(0) \end{array}\right).
\end{equation}
Assuming the qubit initially prepared in a coherent superposition of its dressed states, that is $\cos\theta |A\rangle+\sin\theta\ |B\rangle $, one gets $p_\pm(0)=(1\pm\sin(2\theta))/2$.
Therefore, the quantum probability $p_+(\tau)$ is
\begin{equation} 
p_+(\tau)=\frac{1}{2}\left[1+\frac{1}{2}\sin(2\theta)(A(\tau)+A^{*}(\tau))\right].
\end{equation}

On the other hand, in the presence of a nonselective measurement at time $t=\tau/2$ the (classical) probability is similarly defined at time $\tau$ by \cite{LG9}
\begin{equation}
    \left(\begin{array}{cc} p^{\prime}_+(\tau) \\ p^{\prime}_-(\tau) \end{array}\right)=\Lambda(\tau,\tau/2)\Lambda(\tau/2,0)\left(\begin{array}{cc} p_+(0) \\ p_-(0) \end{array}\right).
\end{equation}
Therefore, the classical probability of finding the open controlled qubit in the state $|+ \rangle $ at time $\tau$ results to be
\begin{equation}
     p'_+(\tau)=\frac{1}{2}\left[1+\frac{1}{4}\sin(2\theta)(A(\tau/2)+A^{*}(\tau/2))^2\right].
\end{equation}

Finally, the quantum witness $W_\mathrm{q}(\tau)=|p_+(\tau)-p'_+(\tau)|$ for our system can be explicitly obtained.


\end{document}